\documentclass[aps,twocolumn,showpacs]{revtex4}
\usepackage{amssymb}
\usepackage{amsmath}
\usepackage{epsfig}
\newcommand{\sech}{\rm sech}
\newcommand{\cn}{\rm cn}

\newcommand{\sn}{\rm sn}
\newcommand{\dn}{\rm dn}
\begin{document}
\title{\bf  Bistability  in sine-Gordon: the ideal switch}
\author {R. Khomeriki${}^{1,2}$,  J. Leon${}^1$}
\affiliation {(${\ }^1$) Laboratoire de Physique Th\'eorique et 
Astroparticules\\
CNRS-UMR5207, Universit\'e Montpellier 2, 34095 Montpellier  (France)\\
 (${\, }^2$) Physics Department, Tbilisi State University,
0128 Tbilisi (Georgia)}

\begin{abstract}  The sine-Gordon equation, used as the representative
nonlinear wave equation, presents a bistable behavior resulting from
nonlinearity and generating hysteresis properties. We show that the process can
be understood in a comprehensive analytical formulation and that it is a
generic property of nonlinear systems possessing a natural band gap. The
approach allows to discover that sine-Gordon can work as an {\it ideal switch}
by reaching a transmissive regime with vanishing driving amplitude.
\end{abstract}

\maketitle

\section{Introduction}

A nonlinear medium submitted to wave irradiation at a frequency in a forbidden
band gap can undergo bistable behavior and present hysteresis properties. This
bistability  has attracted much attention, e.g. in nonlinear optics as a means
for a medium to switch from total reflection to partial (sometimes total)
transmission \cite{winful}, or in superconducting junction devices as a means
to conceive amplifiers that {\it ``remain efficient in the quantum limit''}
\cite{dyn-bif}. 

We attempt to give here a comprehensive interpretation of this phenomenon in
terms of both analytical description and numerical simulations, in order to
unveil a particular stationary regime presenting a non-zero output for
vanishing input, what we call the {\it ideal switch} and which allows for
detection of (almost) vanishing signal. 

To that end we consider the sine-Gordon equation on the finite interval
$x\in[0,L]$
\begin{equation}\label{SG}
u_{tt}-u_{xx}+\sin u =0,\end{equation}
associated to the boundary value problem
\begin{equation}\label{bound}
u(0,t)=f(t),\quad u_x(L,t)=0,
\end{equation}
on a vanishing initial state (with $f(0)=0$ and $f_t(0)=0$ for compatibility of
the initial and boundary values). 

This is a quite standard problem in physics of a wave equation with a forced
extremity (Dirichlet boundary condition in $x=0$) and a free end (Neumann
boundary condition in $x=L$), associated to Cauchy initial data at $t=0$. A
related physical situation is for instance a long Josephson junction 
\cite{agarwal,ustinov} or an array of coupled short junctions (Josephson
superlattice)  \cite{remoiss,olsen}. Note that, depending on  the used external
driving, the boundary \eqref{bound} has possibly to be replaced with
$u_x(0,t)$.

An important subclass of boundary $f(t)$ is  constant amplitude periodic driving
at a frequency in the natural band gap of the system, namely
\begin{equation}\label{periodic}
f(t)=B_0\,\cos (\Omega t),\quad \Omega<1, \end{equation}
after a convenient transient sequence where $f(t)$ grows from a vanishing
amplitude to the value $B_0$ to avoid initial shock. While for a linear system
this boundary excitation does not flow through, nonlinearity allows for energy
transmission above the threshold amplitude which reads (in the semi-infinite
case $L\to\infty$) 
\begin{equation}\label{supra}
B_s=4\arctan (b_s),\quad b_s^2=\frac{1-\Omega^2}{\Omega^2}.\end{equation}
This is  called {\em nonlinear supratransmission} \cite{geniet} and happens by
emission of solitons (moving breathers) that propagate in the nonlinear medium. 

This process, quite  generic, has been experimentally realized on a chain of
coupled pendula \cite{supratrans}, and applies for instance in discrete systems
of coupled waveguide arrays \cite{ramaz} where the forbidden gap results from
discreteness, or else in Bragg media (periodic dielectric structures) under
constant micro-wave irradiation in the photonic band gap \cite{bragg}. In
Josephson junctions arrays, submitted to microwave excitation, the boundary
$u_x(0,t)=f(t)$  induces the threshold  $B_s=2(1-\Omega^2)$
\cite{supratrans}.

In the finite line case considered here, we shall again find a nonlinear
supratransmission threshold which tends to the value \eqref{supra} for large
$L$. But a property far less understood is the hysteresis loop obtained by
decreasing the amplitude excitation $B_0$ from the threshold $B_s$. This
property has been for instance observed on numerical simulations 
\cite{remoiss,olsen} in the context of Josephson superlattices, but both
the analytical expression of the threshold and the very nonlinear mechanism 
involved have not been clarified. 

We shall establish a general procedure to determine the threshold by studying
the standing periodic solutions of sine-Gordon which {\it synchronize} to the
driving frequency $\Omega$ and {\it adapts} to the driving amplitude $B_0$.
Although these two conditions are sufficient to determine {\it completely} the
solution, it is not {\it uniquely} defined. Indeed we shall prove that a fixed
set of physical parameters  $\{L,\ \Omega,\ B_0\}$ may be related to more than
one solution. This is the principle that leads to bistability when $B_0<B_s$.

As an interesting consequence we obtain that there exists a regime where a
vanishing input amplitude $B_0$ produces a non-vanishing output amplitude. This
process shows that sine-Gordon can be thought of as an {\it ideal switch} along
the hysteresis loop from zero to zero input amplitudes.

The paper is organized as follows: in the next section we display the set of
explicit solutions to sine-Gordon on a length $L$ submitted to the only
requirements that the input boundary amplitude be $B_0$ and the period of the
solution be $2\pi/\Omega$. The following section is devoted to the analytical
definition and evaluation of the threshold of bistability. Then we show by
numerical simulations in section 4 that those explicit solution are indeed
produced by the boundary driving \eqref{periodic} and we check bistability
predictions. In particular we compute, for a more realistic damped sine-Gordon
model, the power released by the boundary driver to the medium and find that
after having switched, this power is 2 to 3 orders of magnitude larger than
before the switch.

\section{Explicit solutions}

\subsection{General expressions.}

Under boundary condition \eqref{periodic}, in order to describe the periodic
asymptotic regime reached in numerical simulations, we follow \cite{scott} and 
seek a solution
\begin{equation}\label{arctan}
u(x,t)=4\arctan[X(x)T(t)].\end{equation}
The boundary condition in $u_x(L,t)=0$ then reads
\begin{equation}\label{end}
X'(L)=0,\quad X(L)=A,\end{equation}
where we have defined the amplitude parameter $A$ such as to scale $T(t)$
to unity, in other words
\begin{equation}\label{normal}
\exists t_0\ :\ T'(t_0)=0,\quad T(t_0)=1.\end{equation}
By inserting expression \eqref{arctan} in the sine-Gordon equation \eqref{SG}
and by use of constraints \eqref{end} and \eqref{normal}, we obtain 
differential equations with a unique free parameter $\alpha$:
\begin{align}
& (X')^2=\alpha \Gamma (A^2-X^2)(X^2+\frac1{\Gamma A^2}),\label{eq-X}\\
& (T')^2=\alpha (1-T^2)(T^2+\Gamma),\label{eq-T}
\end{align}
where  prime denotes differentiation  and where
\begin{equation}\Gamma=\frac1{A^2}+\frac1{\alpha (1+A^2)}.\end{equation}
Thanks to \eqref{end} and \eqref{normal} the equation for $X(x)$ is integrated
on  $[x,L]$ and the one for $T(t)$  on $[t,t_0]$ . The solution is then
completely defined  (in terms of elliptic integrals) by the values of the two
parameters  $A$ and $\alpha$,  determined as follows.

Our first fundamental hypothesis is to assume, accordingly with numerical
simulations,  that the solution {\em synchronizes} to the boundary driving,
namely that the function $T(t)$ is periodic with the period of the driver:
\begin{equation}\label{synchro}
T(t+\frac{2\pi}\Omega)=T(t).\end{equation}
The second fundamental hypothesis consists in expressing that the solution 
{\it adapts} to the driving amplitude $B_0=4\arctan(a)$, which gives
\begin{equation}\label{input}
X(0)=a.\end{equation}
The two relations \eqref{synchro} and \eqref{input} constitute a closed system
of equations for the two unknowns $A$ and $\alpha$ in terms of the physical
constants $a$, $\Omega$ and $L$.

The point is that bistability occurs because the solution of
\eqref{eq-X}\eqref{eq-T} drastically depends on the sign of $\alpha$. We shall
indeed discover that there may exist different solutions that do synchronize to
$\Omega$ and adapts to $a$. In other words, for any fixed $\Omega$ and $L$, a
given input amplitude $a$  may correspond to more than one value of the output
amplitude $A$ as depicted on fig.\ref{fig:bifurcation}. 

\subsection{ Type I solutions.}

We call {\it type I solutions} those obtained for $\alpha>0$ (implying $\Gamma
>0$) for which we obtain \cite{byrd}
\begin{align}
&T(t)=\cn(\omega(t-t_0),\nu),\label{T-I}\\
&X(x)=A\ \cn(k(x-L),\mu) ,\label{X-I}\\
&\omega^2=\alpha(1+\Gamma),\quad \nu ^2=\frac1{1+\Gamma},
\nonumber\\
&k^2=\alpha\Gamma (A^2+\frac1{\Gamma A^2}),\quad
\mu ^2=\frac {\Gamma A^4}{1+\Gamma A^4} .
\label{typeI}\end{align}
where $\cn(\cdot,m)$ is the cosine-amplitude Jacobi elliptic function
of modulus $m$.
According to \cite{scott}, the resulting solution $u(x,t)$ is called 
{\it plasma oscillation} and we have from \eqref{typeI}
\begin{equation}\label{disp-I}
\omega^2=k^2+\frac{1-A^2}{1+A^2}.\end{equation}
This relation between the nonlinear wave parameters $\omega$ and $k$ is often
called a {\it nonlinear dispersion relation} but we shall reserve such a
denomination to the true dispersion relation which relates the actual period
$4{\mathbb K}(\mu)/k$ of $X(x)$ to the period $4{\mathbb K}(\nu)/\omega$ 
of $T(t)$.

The parameters $\alpha$ and $A$ are now determined by requiring
synchronization  \eqref{synchro} and input datum \eqref{input}, namely here by
solving for unknowns $\alpha$ and $A$ the system
\begin{equation}\label{typeI-syst}
\Omega\ {\mathbb K}(\nu )=\frac\pi2\ \omega,\quad a=A\ \cn(kL,\mu).
\end{equation}
 (Note: the complete elliptic integral ${\mathbb K}(\nu )$ is well defined as 
for $\Gamma>0$ we have $0<\nu ^2<1$.)

It is useful to express the above solution in terms of the two wave
parameters $\omega$ and $k$. This is done by using \eqref{disp-I} to eliminate
$A$ and $\alpha$ from the definitions of $\mu$ and $\nu$. We get
\begin{align}\label{paramI}
& A^2=\frac {1-\omega^2+k^2}{1+\omega^2-k^2},\nonumber\\
& \nu^2=\frac {k^4-(\omega^2-1)^2}{4\omega^2}, \\
& \mu^2=\frac {4k^2}{(1+k^2)^2-\omega^4}.\nonumber
\end{align}
System \eqref{typeI-syst} appears then as an equation for the determination of
the parameters $\omega$ and $k$ from the data of the length $L$, the boundary
driver's frequency $\Omega$ and amplitude $a$.

\subsection{ Type II solutions.}

New types of solutions are obtained for $\alpha<0$ for which the evolution
\eqref{eq-T} of $T(t)$ requires $\Gamma<0$ in order to guarantee the constraint
\eqref{normal}. Defining then
\begin{equation}
\beta=-\alpha,\quad \Lambda=-\Gamma=\frac1{\beta (1+A^2)}-\frac1{A^2},
\end{equation}
the basic equations \eqref{eq-X} and \eqref{eq-T} become
\begin{align}\label{eq-II-T}
&(T')^2=\beta (1-T^2)(\Lambda-T^2),\\
&(X')^2=\beta \Lambda (A^2-X^2)(X^2-\frac1{\Lambda A^2}).\label{eq-II-X}
\end{align}
It appears that the constraint \eqref{normal} which states that
$T(t_0)=1$, requires $\Lambda>1$, namely
\begin{equation}\label{cond-beta-T}
\beta<\frac{A^2}{(1+A^2)^2},\end{equation}
a condition that must be checked a posteriori when computing $\beta$ from
the synchronization constraint. 

The structure of the equation \eqref{eq-II-X} implies two classes of solutions 
depending on the relative values of $A^2$ and $1/(\Lambda A^2)$.
Type II solutions are obtained for $\Lambda A^4>1$ which, together with
constraint  \eqref{cond-beta-T}, reads 
\begin{align}
&A^2>1\ : \   0<\beta<A^2/(1+A^2)^2,\\
&A^2<1\ : \   0< \beta<A^4/(1+A^2)^2,
\end{align}
The solution of \eqref{eq-II-T}\eqref{eq-II-X} can now be obtained as
\begin{align}
&T(t)=\sn(\omega(t-t_1),\nu),\quad t_1=t_0+{\mathbb K}(\nu)/\omega
\label{T-II}\\
&X(x)=A\ \dn(k(x-L),\mu) ,\label{X-II}\\
&\omega^2=\beta\Lambda,\quad \nu^2=\frac1\Lambda,\nonumber\\
&k^2=\beta\Lambda A^2,\quad \mu^2=1-\frac1{\Lambda A^4},
\label{typeII}\end{align}
and we have the relation
\begin{equation}\label{disp-II}
\omega^2=\frac{k^2}{A^2},\end{equation}
between the wave parameters $k$ and $\omega$ for type II solution.

The parameters $\beta$ and $A$ are determined as before by requiring
synchronization  \eqref{synchro} and input datum \eqref{input}, namely 
\begin{equation}\label{typeII-syst}
\Omega\ {\mathbb K}(\nu )=\frac\pi2\ \omega,\quad a=A\ \dn(kL,\mu).
\end{equation}
Although the above equation, for real valued parameters, has two sets of
solutions $\{\beta,A\}$, only one set verifies the constraint $\Lambda A^4>1$. 

As before, we express the type II solution in terms of the wave
parameters $\omega$ and $k$, by means of \eqref{disp-II} to eliminate
$A$ and $\beta$ in $\mu$ and $\nu$. We obtain
\begin{align}\label{paramII}
& A^2=\frac {k^2}{\omega^2},\nonumber\\
& \nu^2=\frac {k^2}{\omega^2}\, \frac {1-\omega^2-k^2}{\omega^2+k^2}, \\
& \mu^2=1-\frac {\omega^2}{k^2}\, \frac {1-\omega^2-k^2}{\omega^2+k^2}.
\nonumber\end{align}
System \eqref{typeII-syst}  determines then  the parameters $\omega$ and $k$
from the data of  $L$,  $\Omega$ and $a$.

\begin{figure}[ht]
\centerline{\epsfig{file=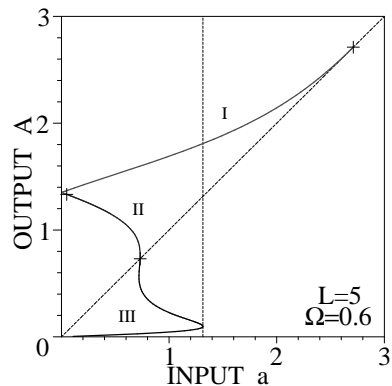,height=5cm,width=5cm}}
\caption {Plot of the curves $|A(a)|$ in the three cases for $L=5$ and
$\Omega=0.6$. The crosses indicate the points where the solution 
changes from one type to the other. The vertical line shows the threshold
amplitude $a_s$ (next section) above which supratransmission occurs by
emission of solitons.}
\label{fig:bifurcation}\end{figure}

\subsection{ Type III solutions.}

The type III solution is obtained still for $\alpha<0$ when $\Lambda A^4<1$.
Such can be realized only in the case $A^2<1$ by requiring
\begin{equation}  
 A^4/(1+A^2)^2<\beta<A^2/(1+A^2)^2.
\end{equation}
The solution of \eqref{eq-II-T}\eqref{eq-II-X} now reads
\begin{align}
&T(t)=\sn(\omega(t-t_1),\nu), \quad t_1=t_0+{\mathbb K}(\nu)/\omega
\label{T-III} \\
&X(x)= A\,\dn^{-1}(k(x-L),\mu) , \label{X-III} \\
&\omega^2=\beta\Lambda,\quad \nu^2=\frac1\Lambda,\quad 
k^2=\frac\beta{ A^2},\quad \mu^2=1-\Lambda A^4,
\label{typeIII}\end{align}
and the wave parameters obey 
\begin{equation}\label{disp-III}
\omega^2=\frac{1}{1+A^2}-k^2.\end{equation}
The synchronization condition  \eqref{synchro} and input datum \eqref{input}
furnish here the system
\begin{equation}\label{typeIII-syst}
\Omega\ {\mathbb K}(\nu )=\frac\pi2\ \omega,\quad 
a=\frac A{\dn(kL,\mu)}.
\end{equation}
for the unknowns parameters $\beta$ and $A$. The same remark as for the type II
solution holds here, namely that the system \eqref{typeIII-syst} has two sets
of solutions $\{\beta,A\}$  but only one verifies the constraint  
$\Lambda A^4<1$. 

Again  the type III solution can be expressed entirely in terms of the two
parameters $\omega$ and $k$ by means of \eqref{disp-III} which gives
\begin{align}\label{paramIII}
& A^2=\frac {1-\omega^2-k^2}{\omega^2+k^2},\nonumber\\
& \nu^2=\frac {k^2}{\omega^2}\, \frac {1-\omega^2-k^2}{\omega^2+k^2}, 
\\
& \mu^2=1-\frac {\omega^2}{k^2}\, \frac {1-\omega^2-k^2}{\omega^2+k^2}.
\nonumber
\end{align}
Then \eqref{typeIII-syst}  is a system for  the parameters $\omega$ and $k$
in terms of the data $L$,  $\Omega$ and $a$.


\section{Bistability thresholds}

The above 3 solutions are now used to describe analytically bistability
properties of sine-Gordon. The first step is to define and calculate, at given
length $L$, the threshold  $a_s$ as functions of the driving frequency
$\Omega$.  Last, decreasing the input amplitude from $a_s$, once a transmission
regime has been reached, the system locks to the type-I solution which holds
down to a vanishing driving amplitude. This is a property that makes sine-Gordon
as the {\it ideal switch} and allows to understand how it can be used to
amplify weak (vanishing) signals.

\subsection{Transmission threshold}

As shown by figure \ref{fig:bifurcation}, increasing the input amplitude from
$a=0$ generates the type III solution. This solution has a maximum input value
$a_s$ resulting from \eqref{typeIII-syst} as the point where $\dn(kL,\mu)$ 
reaches its minimum value $\sqrt{1-\mu^2}$, namely
\begin{equation}
a_s^2=\frac {A^2}{1-\mu^2}.
\end{equation}
Such a definition of the threshold $a_s$ is more conveniently written in terms
of $\omega$ and $k$ through  \eqref{paramIII} as
\begin{equation}\label{as}
a_s^2=\frac {k^2}{\omega^2}.
\end{equation}
In this equation, the parameters $\omega$ and $k$ are determined through 
\eqref{typeIII-syst}, which, at the threshold,  can also be written as the 
system
\begin{equation}\label{threshold-rel}
\Omega\ {\mathbb K}(\nu )=\frac\pi2\ \omega ,\quad kL={\mathbb K}(\mu) ,
\end{equation} 
with $\nu$ and $\mu$ given by \eqref{paramIII}.
\begin{figure}[ht]
\centerline{\epsfig{file=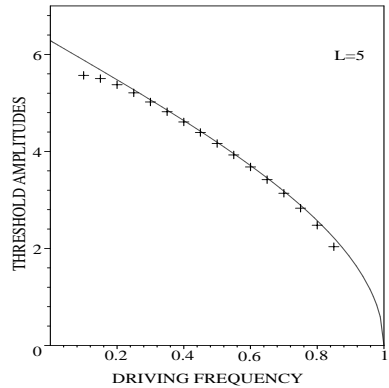,height=5cm,width=5cm}}
\caption {Input amplitude $4\arctan(a_s)$ at the threshold, solution of
\eqref{as}\eqref{threshold-rel} as a function of $\Omega$ for $L=5$ (crosses)
and compared to its limit value  $4\arctan(b_s)$  as $L\to\infty$, given by  
\eqref{lim-supra} (full line).}
\label{fig:thresholds}
\end{figure}

The figure \ref{fig:thresholds} shows that the threshold $a_s$ is quite close to
the expression of $b_s$ in \eqref{supra}.  This property is demonstrated in
general by studying the limit $L\to\infty$ in the type III solution. At
threshold amplitude, the relation \eqref{threshold-rel} between $L$ and $\mu$
gives the necessary condition
\begin{equation}\label{lim-disp}
L\to\infty\ \Rightarrow\ \mu\to 1\ \Rightarrow\ \omega^2\to 1-k^2,
\end{equation}
with which the synchronization condition provides 
\begin{equation}
L\to\infty\ \Rightarrow\ \nu\to 0\ \Rightarrow\ \omega\to\Omega.
\end{equation}
With this in hands, the expression \eqref{as} of the threshold
readily
gives the expression \eqref{supra}, namely
\begin{equation}\label{lim-supra}
L\to\infty\ \Rightarrow\ a_s^2\to b_s^2=\frac{1-\Omega^2}{\Omega^2}.
\end{equation}

{\em Remark:} it is instructive to compute also the limit $L\to\infty$ on the
solution itself at the threshold where from \eqref{paramIII} and
\eqref{lim-disp} obviously $A\to0$. In that case we rewrite the solution $X(x)$
of \eqref{X-III} as
\begin{equation}
X(x)= \frac {a\,\dn(kL,\mu)}{\dn(k(x-L),\mu)},\end{equation}
expand the denominator, take the limit $L\to\infty$ first and make then
$\mu\to1$. We obtain
$X(x)$ of \eqref{X-III} as
\begin{equation}
X(x)\underset{L\to\infty}{\longrightarrow} a_s\,\sech (kx).
\end{equation}
The same procedure applied to $T(t)$ in \eqref{T-III}, with $\nu\to0$, provides
\begin{equation}
T(t)\underset{L\to\infty}{\longrightarrow}\cos(\omega(t-t_0)),
\end{equation}
\begin{figure}[t]
\centerline{\epsfig{file=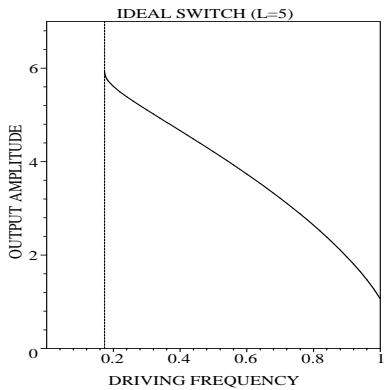,width=5cm,height=5cm}}
\caption{Output amplitude $4\arctan(A)$  when $a=0$ as a function of
$\Omega$ for $L=5$ obtained from \eqref{ideal-A} by solving \eqref{ideal-sys}.}
\label{fig:switch}\end{figure}
and the resulting solution $u(x,t)$ of sine-Gordon on the semi-line is the
stationnary breather centered in $x=0$ accordingly with \cite{geniet}.

\subsection{Ideal switch}

Considering the type I solution, expression \eqref{typeI-syst} that links the
input $a$ to the output $A$ can produce $a=0$ with $A\ne0$, such as to generate
a regime of non-vanishing output value with a  vanishing input amplitude,
the {\it ideal switching} regime. This is the case when
\begin{equation}\label{ideal-sys}
kL={\mathbb K}(\mu),\quad \Omega\ {\mathbb K}(\nu )=\frac\pi2\ \omega.
\end{equation}
where $A$, $\mu$ and $\nu$ are related to $\omega$ and $k$ through
\eqref{paramI}. Note the formal analogy with \eqref{threshold-rel} where the
parameters $A$, $\mu$ and $\nu$ are different.
\begin{figure}[hb] \centerline
{\epsfig{file=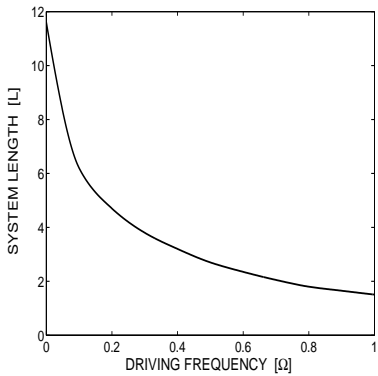,width=5cm,height=5cm}} \caption {Minimal length of
the system below which one  does not have ideal switch  (the solution of
\eqref{ideal-sys} ceases to exist).} \label{fig:switch-L}\end{figure}

This is a system of equations for $\{\omega,k\}$ whose solution then produces
the seeked output amplitude by
\begin{equation}\label{ideal-A}
A^2=\frac {1-\omega^2+k^2}{1+\omega^2-k^2},
\end{equation}
defined by \eqref{ideal-sys} in terms of the physical entries $L$ and $\Omega$.
We have plotted in fig.\ref{fig:switch} the output amplitude  
$4\arctan A$  in the ideal switching case ($a=0$) as a function of $\Omega$
for length $L=5$.

We observe that, for a given length, there exists a threshold in frequency below
which no ideal switching is allowed. This is understood by observing that $A^2$
diverges when $\omega^2\to k^2-1$, for which $\mu\to 1$ (a limit threshold that
has exactly the same origin as in \eqref{lim-disp} when $L\to\infty$).
Conversely, at given driver frequency $\Omega$, there exists a minimum length
$L$ of the medium to obtain an ideal switch, it is displayed on fig.
\ref{fig:switch-L}.

Let us remark  that the notion of {\it nonlinear dispersion relation} is not
useful to predict, at given driving frequency $\Omega$, the minimum driver
amplitude that generates transmission, as indeed we have here an example where
this minimum is simply vanishing.

\section{Numerical simulations}

\subsection{Damping and boundary driving.}

Bistable properties, and in particular ideal switching, have been analytically
described in the integrable case  \eqref{SG}. However, any realistic physical
situation must take into account the damping inherent to the medium.
The simplest way to include damping is to assume the model
\begin{equation}
u_{tt}+\gamma u_t-u_{xx}+\sin u =0,\label{SG-damped}
\end{equation}
associated with the initial-boundary value problem \eqref{bound} in the
particular subclass  \eqref{periodic}. We study here the bistable properties of
\eqref{SG-damped} by numerical simulations and compare the results to the
analytical predictions.
\begin{figure}[hb]
\centerline{\epsfig{file=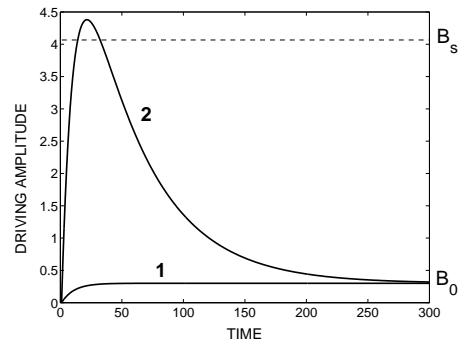,width=0.7\linewidth}}
\caption{Two different paths for driving the sine-Gordon system. $B_0$ is the 
driving amplitude in stationary regime and $B_s$ is a supratransmission 
threshold.}\label{fig:driving}\end{figure}

To that end the system is driven at the input boundary $x=0$ with a bandgap
frequency and time dependent amplitude as follows: in a first numerical
simulation, the amplitude is smoothly increased up to the value $B_0$ smaller
than the supratransmission threshold $B_s$. In a second simulation, the
amplitude is increased up to a value exceeding the supratransmission threshold
and then, after a time sufficient to generate moving breathers, it is decreased
to the same value $B_0$ as in the first case. The figure \ref{fig:driving}
displays the two time variations of the driving amplitude that we have used in
the numerical simulations. After having reached a stationary regime, although
the driving amplitudes $B_0$ are equal in both cases, the dynamics drastically
differ in those two cases as it is expected from the analytical consideration
presented above and described hereafter.

\subsection{Evidence of bistability.}

\begin{figure}[t]
\centerline {\epsfig{file=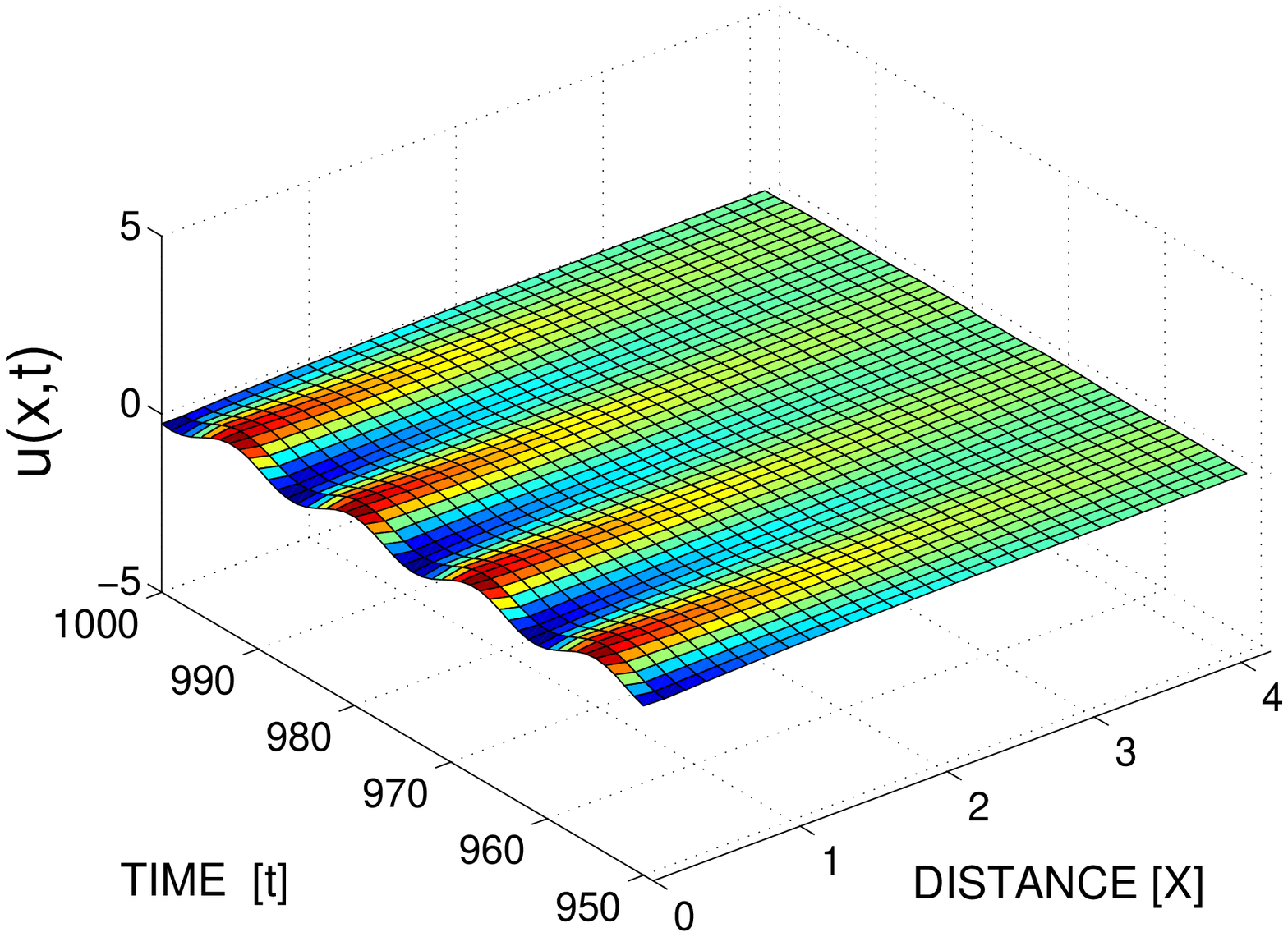,width=0.8\linewidth}}
\centerline {\epsfig{file=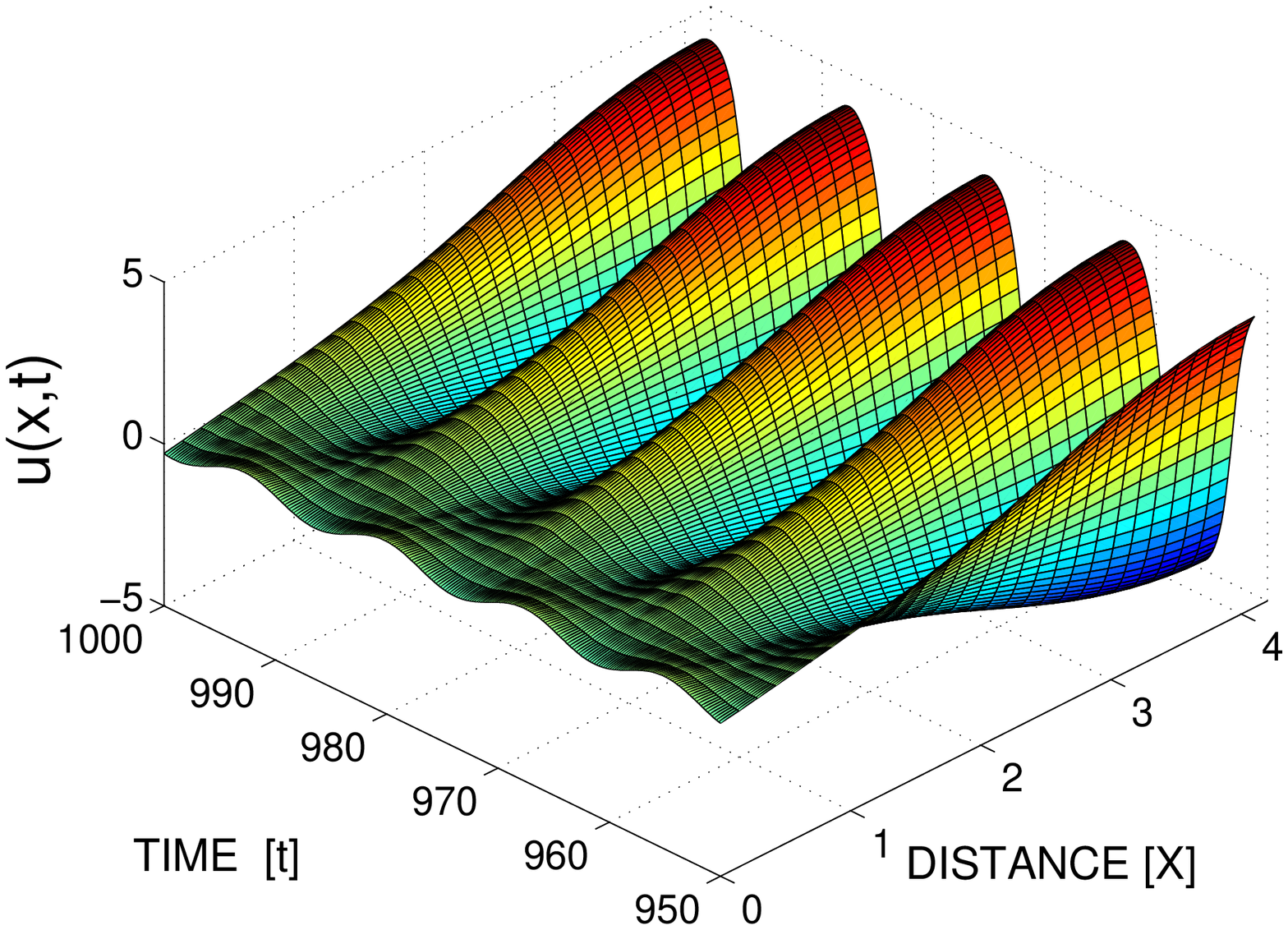,width=0.8\linewidth}}
\caption {Three dimensional plots of the sine-Gordon system dynamics after
having reached stationary regime.  Upper plot corresponds to the driving path 1
of fig.\ref{fig:driving} and lower plot to path 2. In both cases driving
frequency is $\Omega=0.5$, stationary  driving amplitude $B_0=0.3$, damping
coefficient $\gamma=0.01$ and system length $L=4.1$.}
\label{fig:surf}\end{figure}

In most of our numerical simulations we choose the driving frequency in the
middle of the band gap $\Omega=0.5$, use damping parameter $\gamma=0.01$ and
length $L=4.1$. For the driving amplitude along path 1 in fig.\ref{fig:driving},
we always observe a stationary regime with decaying profile of the standing
wave, very well described by the exact analytical solution of type III
(\ref{T-III})(\ref{X-III}). Instead, when  driving along path 2, we get the
picture corresponding to the exact type I solution (\ref{T-I})(\ref{X-I}). Those
two drastically different behaviors of the system are displayed as three
dimensional plots in fig.\ref{fig:surf}.

It is remarkable on the second picture of fig.\ref{fig:surf} that the system
has locked to a stationary solution with a small driving amplitude (here
$B_0=0.3$) and a large output amplitude (evaluated at $B=4.35$). Let us mention
that we can drive the system with  amplitudes down to $B_0=0.1$ and still have
the type I solution (large output amplitude) despite  presence of
damping in the system, getting thus a regime of almost ideal switch, or almost
perfect  detector.

Another remark is that the system never locks to the exact solution of type II
(\ref{T-II})(\ref{X-II}), simply because this solution is unstable. For
instance, if that exact solution is used as an initial condition in
sine-Gordon, it eventually breaks down.
\begin{figure}[t]
\centerline {\epsfig{file=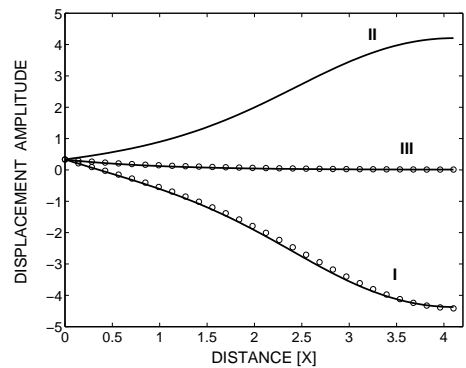,width=0.7\linewidth}}
\caption {Comparison of analytical expressions (solid lines) for the profiles of
standing waves \eqref{X-III}, \eqref{X-II}, \eqref{X-I}, corresponding to
solutions of type III, II and I  respectively, and numerical simulations
(circles). The type III solution is reached along path 1 of fig.
\ref{fig:driving} while the type I results from path 2. The type II solution,
unstable, is never reproduced by the system.\label{fig:analnumer}}
\end{figure}
\begin{figure}[b]
\centerline{\epsfig{file=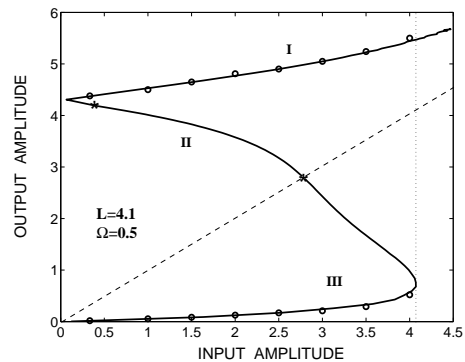,width=0.7\linewidth}}
\caption {Comparison of input-output amplitude dependence obtained from
numerical  simulations (circles) with analytical curves derived from formulas 
\eqref{typeI-syst}, \eqref{typeII-syst}, \eqref{typeIII-syst}.\label{fig:fit}}
\end{figure}

To be complete, we compare the analytical expressions of the standing waves
profiles of  \eqref{X-III}, \eqref{X-II} and \eqref{X-I} to the results of
numerical simulations in the case of the two different driving paths in the
context of fig.\ref{fig:surf}. The result is displayed in figure
\ref{fig:analnumer} where the obtained perfect matching shows that indeed the
system locks to the analytical solution obtained by assuming frequency
synchronization and amplitude matching. This is completed by plotting in
fig.\ref{fig:fit} the input-output amplitude dependence obtained from numerical
simulations and its comparison with analytical curves derived from formulas
\eqref{typeI-syst}, \eqref{typeII-syst} and \eqref{typeIII-syst}.

\subsection{Energetic considerations.}

The physically useful bistable nature of the system manifests in large
difference between the energy dissipation in the two stationary regimes. The
averaged energy released from the driver in unit time, can be expressed as
\begin{equation}
P=\frac{\Omega}{2\pi}\int\limits_0^{2\pi/\Omega} dt\int\limits_0^L 
dx\ \gamma\,u_t^2. \label{power}
\end{equation}
Indeed one can easily obtain
\begin{align}
&\frac{\partial}{\partial t}\int\limits_0^Ldx\,{\cal H}=-\int\limits_0^L 
dx\,\gamma\,u_t^2-\left.(u_xu_t)\right|_{x=0},\nonumber\\
&{\cal H}=\frac12u_t^2+\frac12u_x^2+1-\cos u,
\end{align}
which by averaging on one period furnishes \eqref{power} as the boundary value
in $x=0$ is periodic.
\begin{figure}[t]
\centerline{\epsfig{file=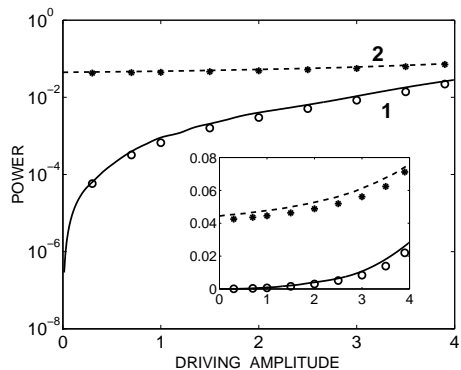,width=0.7\linewidth}}
\caption { Power $P$ of \eqref{power}  in terms of the driving amplitude in
lin-log (main plot) and lin-lin (inset) scales. Solid line (solution
(\ref{T-I},\ref{X-I})) and circles (simulations) correspond to driving regime 1
of fig.\ref{fig:driving}. Dashed line (solution (\ref{T-III},\ref{X-III})) and
asterisks (simulations)  result from driving regime 2. Length is  $L=4.1$,
driving frequency $\Omega=0.5$ and damping coefficient $\gamma=0.01$.}
\label{fig:lin-log}\end{figure}

Although the analytical solution of the sine-Gordon equation \eqref{SG} are not
solutions of the damped version \eqref{SG-damped}, we may compare the power
defined above obtained by numerical simulations of \eqref{SG-damped} to the
expression \eqref{power} where $u$ is simply replaced by the exact solution
(type III before the switch, type I after). The result of this comparison is
displayed in fig.\ref{fig:lin-log} wher we see that expression  \eqref{power}
with analytical solutions fit strikingly well the numerical simulations, and
that the power $P$ after the switch is  two to three orders of magnitude
greater than before the switch.

\section{Comments and conclusion.}

It is worth mentioning that while the analytical solution of type I holds for
any length $L$, in a realistic physical system (nonzero damping), the situation
is different. In particular, for large L and low driving amplitudes, when
several nodes of type I standing wave solution are present, the solution cannot
survive and decays to type III solution. This is understood by the following
simple argument: all of the exact solutions derived in the previous sections
are standing waves, i.e. they do not generate  energy flux. Thus, regions of
the system far from the boundary  cannot gain energy from the driver and the
oscillations  will  eventually fade away. 

We have essentially demonstrated, both by analytical and numerical treatment,
that the bistable property on the sine-Gordon system allows to generate a
particular regime that works as an ideal switch: nonzero output for vanishing
input. In a realistic physical system (including damping) the property is
conserved but for small (non-vanishing) input. This nonlinear hysteretis has
been shown to correspond to quite determinant differences in the power released
by the driver to the system.

{\bf acknowledgements:}  It is a pleasure to thank Yu.S. Kivshar for the useful
information regarding the recent experiments on bifurcations in Josephson
junctions.  One of us (R.Kh.) aknowledge support of the France-NATO visiting
scientist fellowship award.

\end{document}